\documentclass[aps,prl,twocolumn,tworchcolumn, showpacs,preprintnumbers,amsmath,amssymb,floatfix]{revtex4}
\usepackage{graphicx}% Include figure files
\usepackage[usenames]{color}
\renewcommand{\vec}[1]{\mathbf{#1}}

\begin{document}

\title{A method of state-selective transfer of atoms between microtraps based on the Franck-Condon Principle.}

\author{A. B. Deb, G. Smirne, R. M. Godun, C. J. Foot}
\affiliation{Clarendon Laboratory, University of Oxford, Parks Road,
Oxford, OX1 3PU, UK}
%\author{G. Smirne}
%\affiliation{Clarendon Laboratory, University of Oxford, Parks Road,
%Oxford, OX1 3PU, UK}

%\author{R. M. Godun}
%\affiliation{Clarendon Laboratory, University of Oxford, Parks Road,
%Oxford, OX1 3PU, UK}
%\author{C. J. Foot}
%\affiliation{Clarendon Laboratory, University of Oxford, Parks Road,
%Oxford, OX1 3PU, UK}

\date{\today}

\begin{abstract}
%BEC, time-varying potential in 2D, arbitrary pattern, ferro-electric, fast,
%reconfigurable phase grating, transport and splitting.

We present a method of transferring a cold atom between spatially
separated microtraps by means of a Raman transition between the
ground motional states of the two traps. The intermediate states for
the Raman transition are the vibrational levels of a third
microtrap, and we determine the experimental conditions for which
the overlap of the wave functions leads to an efficient transfer.
There is a close analogy with the Franck-Condon principle in the
spectroscopy of molecules. Spin-dependent manipulation of neutral
atoms in microtraps has important applications in quantum
information processing. We also show that starting with several
atoms, precisely one atom can be transferred to the final potential
well hence giving deterministic preparation of single atoms.
\end{abstract}

%\pacs{32.80.Pj, 42.40.My, 03.75.Kk}

\maketitle

%The ability to move atomic wavefunctions, make them interact in a
%controlled spin-dependent way and separate them afterwards is the
%key to applications of ultracold atoms in microtraps such as
%creating quantum-entangled states of atoms and for realizing
%collisional gates for quantum computing. A method of achieving this
%has been demonstrated experimentally~\cite{mandel2003} in which the
%microtraps were part of a spin-dependent optical lattice that was
%physically moved to transport the atomic wavepackets though space.
%In this scheme, the trapping light must be relatively close to the
%atomic resonance for the spin-dependence and this introduces
%decoherence due to inelastic light scattering and also motional
%heating. Spin-dependent potentials are introduced in the same way in
%the scheme described here but the atoms are only exposed to this
%radiation transiently - in the initial and final states the atoms
%are stored in traps where scattering rate is low. Instead of dipole
%traps created by laser light, these ideas can be implemented using
%strong spin-dependent potentials created in the region close to the
%current-carrying wires on magnetic atom chips; these can be either
%magnetostatic, RF or microwave fields. Beam-splitting, spatial
%separation and recombination of free atomic wavefunctions is the
%basis of atomic interferometry~\cite{Godun2001} and the scheme
%presented here implements this with micro-trapped atoms.

Ultracold atoms in arrays of microtraps have been proposed as
systems for storing quantum information in the internal states of
the atoms \cite{jaksch1999}. To carry out quantum information
processing of the information in such a register, the atoms must
interact in a way that depends on their internal state, but because
the interactions between neutral atoms are short range, this
requires bringing the atomic wavepackets into `contact', i.e.\
putting atoms into the same well. A method of achieving this has
been demonstrated experimentally ~\cite{mandel2003} in which the
microtraps belonged to a spin-dependent optical lattice that was
physically moved to transport the atomic wavepackets though
space---in order to give a spin-dependent potential in that scheme
the trapping light has a frequency closer to the atomic resonance
frequency than would otherwise be the case and this introduced some
undesirable inelastic light scattering leading to decoherence. Also,
the movement of the potentials must be carried out sufficiently
slowly to avoid motional heating. In this paper we analyse a scheme
that differs from~\cite{mandel2003} in the following ways: a) it
does not require the potentials to change position but only to be
turned on and off adiabatically (by controlling the intensity of the
light, which is necessary to loading the trap in any case), and b)
although spin-dependence is introduced in the same way as
in~\cite{mandel2003} the atoms are only exposed to this radiation
transiently; atoms in the initial and final states are stored in
traps where the scattering rate is very low
 and wavepackets are transferred between microtraps by a Raman process. Similar schemes
are implicitly assumed in proposals for creating artificial magnetic
fields \cite{JakschZoller} and fractional occupation of fermions in
an optical lattice \cite{janne1} but here we give explicit values
for realistic experimental conditions. Instead of using dipole traps
created by laser light, these ideas could be implemented using the
strong spin-dependent potentials close to current-carrying wires on
magnetic atom chips; suitable potentials can be either
magnetostatic, associated with RF/microwave radiation, or some
combination of these. To date, however, the distance scale over
which the potentials are modulated by atom chips is considerably
greater than the sub-micron length scales in optical lattices. In
addition to realizing a collisional gate for quantum computation,
the scheme described here can perform the operations of
beam-splitting, spatial separation and recombination of atomic
wavepackets that form the basis of atomic
interferometry~\cite{Godun2001}.

\begin{figure}[h]
\includegraphics[ width=.90\linewidth]{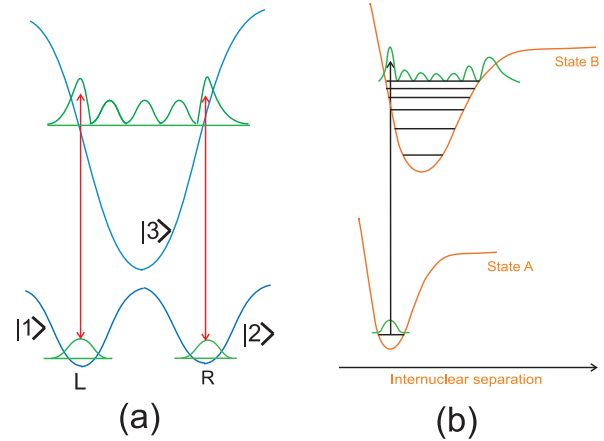}
\caption{\label{fig:FC}(a) Franck-Condon Transfer: The spatial
transfer of neutral atoms between two wells, L and R using the
overlap of vibrational wavefunctions. Each of the vertical
double-headed arrows denotes a microwave transition between
hyperfine levels of the atomic ground configuration (or a Raman
transition between these states). Note that the Franck-Condon
Transfer scheme shown in (a) does not work for electronic
transitions, as explained in this paper. (b) The Franck-Condon
principle explains the relative strengths of the spectral lines
between vibraional levels that arise in the electronic transitions
in diatomic molecules. The strongest line in the vibrational spectra
is indicated by the vertical arrow.}
\end{figure}

In this paper we analyse a scheme based on a Raman transition of a
trapped neutral atom from the ground vibrational state of one well
to that of a neighboring well as shown in Fig.~\ref{fig:FC}(a). The
left and right wells represent the (one-dimensional) potential
experienced by the atom in internal states $|1\rangle$ and
$|2\rangle$ respectively. In the deep central well the atom is in
internal state $|3\rangle$  We assume that the atom is initially in
the lowest vibrational level of the left well. A Raman transition
brings the atom from $|1\rangle$ to $|2\rangle$ where the initial
and final vibrational wave functions have a significant overlap with
the vibrational wave functions of the intermediate state. The scheme
has an analogy with the well-known Franck-Condon principle in
molecular physics. Underlying this principle is the Born-Oppenheimer
approximation which allows the electronic and the nuclear
 motions to be separated and the molecular wave function to be written as a product
 of the electronic wave function and the vibrational wave function:

%\begin{figure}[h]
%\includegraphics[width=.45\linewidth]{molecular1}
%\caption{\label{fig:mol}The Franck-Condon principle explains the
%relative strengths of the spectral lines between vibraional levels
%that arise in the electronic transitions. In this figure, the
%strongest line in the vibrational spectra is marked by the straight
%line X.}
%\end{figure}

 \begin{equation}\label{BO}
    \Psi(\bf{r},\bf{R}) =
    \psi_{e}(\bf{r,R})\psi_{v}(\bf{R}),
 \end{equation}
 where $\bf{r}$ and $\bf{R}$ are the
 electronic and the center-of-mass coordinates, respectively. Thus
 when the molecule undergoes an electric dipole transition from a
 state A to an excited state B (Fig 1(b)), the electric dipole matrix
 element is proportional to
\begin{figure}[h]
\includegraphics[width=0.5\linewidth]{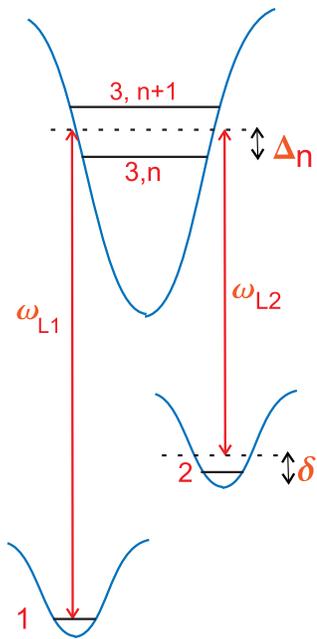}
\caption{\label{fig:three} A schematic of the three potential wells
showing their vibrational levels. Transitions are driven by
radiation at angular frequencies $\omega_{L1}$ and $\omega_{L2}$.
The frequency detunings from resonance $\Delta_n$ and $\delta$ are
indicated.}
\end{figure}
\begin{eqnarray}\label{FC1}\nonumber
    &&\langle\Psi_{A}|(\bf{r} + \bf{R})|\Psi_{B}\rangle \\ \nonumber
    &=& \langle\psi_{e,A}\psi_{v,A}|(\bf{r} + \bf{R})|\psi_{e,A}\psi_{v,B}\rangle\\
    &=& \langle\psi_{v,A}|\psi_{v,B}\rangle\langle\psi_{e,A}|\bf{r}|\psi_{e,B}\rangle
 \end{eqnarray}

  This follows from the orthogonality of the electronic
 eigenstates. (The \emph{vibrational}  eigenstates corresponding to different
 potential functions do not need to be orthogonal). The
 modulus-squared of the Franck-Condon factor
 $\langle\psi_{v,A}|\psi_{v,B}\rangle$ determines the relative strength
 of transitions. Our scheme employs as the internal states
hyperfine states of the atomic ground configuration. The hyperfine
levels have typical energy splitting of order $1$\,GHz, whereas the
vibrational states of dipole traps have energies of tens of kHz, so
we can write the wavefunction as a product of the internal and
vibrational wavefunctions.

%Spatial re-orientation using the overlap of vibrational wavefunction
%is well-known in molecular physics in the context of Franck-Condon
%principle, which states that the strongest line in the vibrational
%spectra is the one that corresponds to the largest overlap of
%initial and final vibrational wavefunctions (Franck-Condon factor).
%Lasers tuned as in Fig.~\ref{fig:mol} transfers with greatest
%probability the molecule to a state of larger average internuclear
%separation. In our case, the strength of the spatial transfer from
%the left well to the right well (Fig.~\ref{fig:FC}) two such overlap
%factors. Therefore we call the transfer a Franck-Condon transfer.

Usually, the potential corresponding to the intermediate atomic
state will have a number of vibrational states with non-zero overlap
with the initial and final states. The alternating parity of
eigenfunctions of a given Hamiltonian leads to the fact that the
overlaps between two nearest vibrational levels tend to cancel each
other (the implications of this fact will be discussed later). We
derive below under what conditions the problem reduces to a
three-level problem. Fig.~\ref{fig:three} shows the situation
schematically. The ground states of the left and the right well,
denoted by $1$ and $2$, are coupled to the vibrational manifold
(denoted by $n$) of the potential associated with the intermediate
atomic state $3$ (these states are denoted by ${3,n}$) by two lasers
or microwaves of frequencies $\omega_{L1}$ and $\omega_{L2}$
respectively.  The Hamiltonian

\begin{equation}
     H = H_{0} + H_{I}
\end{equation}
can be written in the basis $|1\rangle$,$|2\rangle$,
$|3,1\rangle$,$|3,2\rangle$..., as
\begin{equation*}
    =\hbar\begin{pmatrix}
0 & 0 & M_{31,1}& M_{32,1}&M_{33,1}&.&.\\
0 & \omega_2 & M_{31,2} & M_{32,2}&M_{33,2}&.&. \\
M_{31,1}^{*}& M_{31,2}^{*} & \omega_{31}& 0 & 0&.&.\\
M_{32,1}^{*}& M_{32,2}^{*} & 0& \omega_{32} & 0&.&.\\
M_{33,1}^{*}& M_{33,2}^{*} & 0& 0 & \omega_{33}&.&.\\
.&.&.&.&.&.&.\\
.&.&.&.&.&.&.
\end{pmatrix}
\end{equation*}
, where $H_{0}$ is the atomic Hamiltonian, and the interaction
Hamiltonian $H_I$ is given by
 \begin{equation*}
    H_{I,j} = -\vec{d}.\vec{E}_{0,Lj} \hspace{0.1 in},
\end{equation*}
for optical (electric dipole) transitions and
\begin{equation*}
     H_{I,j} = -\vec{\mu}.\vec{B}_{0,Lj}\hspace{0.1 in},
\end{equation*}
for microwave (magnetic dipole) transitions.
\begin{equation*}
    M_{3n,j} = \Omega_{3n,j}\cos(\omega_{Lj}t)
\end{equation*}
Here, $\Omega_{3n,j}$ is the Rabi frequency of the transition
between
  levels $|3,n\rangle$ and $j$ ($=1,2$) which gives the coupling
  strength between these levels induced by the lasers/microwaves
  $\omega_{Lj}$ and is given by,
  \begin{eqnarray}\label{rabii}
   \Omega_{3n,j} &=& \frac{\langle 3n|H_{I,j}|j\rangle}{\hbar}\\
  &=& \frac{\langle 3|H_{I}|j\rangle\langle n|n_{j}\rangle}{\hbar}\\
  &=& ~\widetilde{\Omega}_{3n,j}\sqrt{f_{n,j}}\hspace{0.1 in},
\end{eqnarray}
where $n_j = n_1$ or $n_2$ are the vibrational quantum number of the
initial/final well and n is the vibrational quantum number of the
intermediate well. $\tilde{\Omega}_{3n,j}$ is the Rabi frequency if
the overlap is perfect and $f_{n,j}$ is a dimensionless quantity
representing the overlap of wave function squared. The zero of
energy is taken to be the energy of  $|1, n_{1}=1\rangle$.
Non-resonant transition between internal states $|1\rangle$ and
$|2\rangle$ is ignored.

We make the usual transformation to the interaction picture using
the unitary operator $U = e^{-\frac{i}{\hbar}H_{0}t}$,we obtain the
transformed Hamiltonian,
%\begin{equation*}
    %U = e^{-\frac{i}{\hbar}H_{0}t}\\
    %= \begin{pmatrix}
    %1 & 0 & 0& 0&0&.&.\\
%0 & e^{-i\omega_{2}t} & 0 & 0 & 0&.&. \\
%0& 0 & e^{-i\omega_{31}t}& 0 & 0&.&.\\
%0&  0 & 0& e^{-i\omega_{32}t} & 0&.&.\\
%0& 0 & 0& 0 & e^{-i\omega_{33}t}&.&.\\
%.&.&.&.&.&.&.\\
%.&.&.&.&.&.&.

   % \end{pmatrix}
%\end{equation*}

\begin{equation*}
     \widetilde{H}= \hbar \begin{pmatrix}
0& 0 & N_{31,1}& N_{32,1}&N_{33,1}&.&.\\
0 & 0 & K_{31,2} & K_{32,2} & K_{33,2}&.&. \\
N_{31,1}^{*}& K_{31,2}^{*}& 0& 0 & 0&.&.\\
N_{32,1}^{*}&  K_{32,2}^{*} & 0& 0 & 0&.&.\\
N_{33,1}^{*}& K_{33,2}^{*} & 0& 0 & 0&.&.\\
.&.&.&.&.&.&.\\
.&.&.&.&.&.&.
\end{pmatrix},
\end{equation*}
In the RWA, the matrix elements are
\begin{equation*}
    N_{3n,j} = \frac{1}{2}\Omega_{3n,j}e^{i\Delta_{n}t}
\end{equation*}
\begin{equation*}
    K_{3n,j} = \frac{1}{2}\Omega_{3n,j}e^{i(\Delta_{n}-\delta)t}
\end{equation*}
The frequency detunings are (see Fig.$2$),
\begin{equation*}
    \Delta_{n} = \omega_{L1} - \omega_{3n}
\end{equation*}
and
\begin{equation*}
   \delta = (\omega_{L1} - \omega_{L2}) - (\omega_{2}-\omega_{1}).
\end{equation*}

The time-evolution of the wavefunction
\begin{equation*}
    \widetilde{\Psi}(t) = c_1(t)|1, n_1 =1\rangle + c_2(t)|2, n_2 =1\rangle
    + \displaystyle\sum_{n}c_{3n}(t)|3, n\rangle
\end{equation*}
is given by
\begin{equation*}
    i\hbar\frac{d}{dt}\widetilde{\Psi}(t) =
    \widetilde{H}(t)\widetilde{\Psi}(t).
\end{equation*}

Thereby we obtain a set of coupled first-order differential
equations:

\begin{eqnarray}
 \frac{dc_1}{dt} &=&   i\displaystyle\sum_{n}N_{3n,1}c_{3n}  \\
 \frac{dc_2}{dt} &=&   i\displaystyle\sum_{n}K_{3n,1}c_{3n} \\
 \frac{dc_{3n}}{dt} &=&
    i(N_{3n,1}^{*}c_1 + K_{3n,2}^{*}c_2),\\
    \nonumber
    \end{eqnarray}
for all $n$. We can see from Eq. ($7$), ($8$) and ($9$) that under
the condition:\begin{equation}\label{adia}
    \Delta_{n} \gg \Omega_{3n,1}, \Omega_{3n,2},
\end{equation}
the variables $c_{3n}$ (for all $n$) oscillate very fast compared to
the slow variables $c_1$ and $c_2$ so that `adiabatic elimination'
is possible where we integrate ($9$) considering $c_1$ and $c_2$ as
constants to obtain:
\begin{equation}\label{slow}
    c_{3n} = \frac{\Omega_{3n,1}e^{-i\Delta_{n}t}}{2\Delta_{n}}c_{1} +
    \frac{\Omega_{3n,2}e^{-i(\Delta_{n}-\delta)t}}{2(\Delta_{n}-\delta)}c_{2},
\end{equation}
for all $n$. Substituting in equations ($7$) and ($8$), we obtain
\begin{equation}\label{}
    \frac{dc_1}{dt} =
    i c_1\displaystyle\sum_{n}\frac{|\Omega_{3n,1}|^2}{2\Delta_{n}}
    +
    i c_2\displaystyle\sum_{n}\frac{\Omega_{3n,1}\Omega_{3n,2}e^{i\delta
    t}}{2(\Delta_n
    - \delta)}
\end{equation}
\begin{equation}\label{}
    \frac{dc_2}{dt} =
     i c_1\displaystyle\sum_{n}\frac{\Omega_{3n,1}\Omega_{3n,2}e^{-i\delta
    t}}{2(\Delta_n -
     \delta)} + i c_2\displaystyle\sum_{n}\frac{|\Omega_{3n,2}|^2}{2\Delta_{n}}
\end{equation}
Thus under condition ($10$) the system reduces to an effective
two-level system. The additional condition
\begin{equation}\label{}
    \Delta_{n} \gg \delta,
\end{equation}
 for all $n$ allows us to make another unitary transformation using the operator
\begin{equation*}
    U^{'} = \begin{pmatrix}
    e^{-i\delta t/2} & 0 \\
    0 & e^{i\delta t/2}
    \end{pmatrix}
\end{equation*}
which reduces everything to an effective two-level problem with an
effective Rabi frequency:
\begin{eqnarray}\label{effective1}
    \Omega_{eff} &=&
    \displaystyle\sum_{n}\frac{\Omega_{3n,1}\Omega_{3n,2}}{2\Delta_n}\\
    &=&~ \widetilde{\Omega}_{3,1}\widetilde{\Omega}_{3,2}\displaystyle\sum_{n}\frac{\langle n_1=1|n\rangle\langle n|n_2=1\rangle}{2\Delta_n},
\end{eqnarray}
The effective detuning of the two-level system is given by
\begin{equation}\label{detuning}
    \delta_{eff} ~=~ \delta -
    \displaystyle\sum_{n}(\frac{|\Omega_{3n,1}|^2}{4\Delta_{n}} -
    \frac{|\Omega_{3n,2}|^2}{4\Delta_{n}}).
\end{equation}

$\widetilde{\Omega}_{3,1}$ and $\widetilde{\Omega}_{3,2}$ are
defined in Eq.($6$). They involve matrix elements only pertaining to
the internal states of the atom and the power of the laser or
microwave fields at frequencies $\omega_{L1}$ and $\omega_{L2}$
respectively. The presence of a vibrational manifold is manifested
in the summation. This effective detuning equals the normal detuning
minus the net light shift of the initial and final state in the
presence of a number of intermediate states. For $\delta_{eff} = 0$,
%\begin{equation*}
    %\delta ~=~
    %\displaystyle\sum_{j}(\frac{|\Omega_{3j,1}|^2}{4\Delta_{j}} -
    %\frac{|\Omega_{3j,2}|^2}{4\Delta_{j}}).
%\end{equation*}
 a pulse of length
\begin{equation}\label{pipulse}
   t_{\pi} = \frac{\pi}{\Omega_{eff}}
\end{equation}
transfers an atom initially in the lowest vibrational level of the
left well completely to that of the right well when the conditions
leading to Eq.($16$) are satisfied.

In what follows we discuss the transfer scheme only in the
situations where the initial, the intermediate and the final
internal states are ground hyperfine states of the atom. In the
appendix, we show that the Franck-Condon transfer scheme cannot be
carried out using electronic transitions.

To create potential landscapes that depend on the internal state for
ground hyperfine states of alkali atoms, there are different viable
options -  among them are i) using spin-dependent dipole potentials
created by off-resonant laser light ~\cite{mandel2003}, ii) magnetic
potentials on atoms chips, including magnetostatic trapping modified
by RF fields or near-field microwave potentials
~\cite{Treutlein2005}, or combinations of these or a combination of
any of these with conservative optical dipole potentials. As shown
below, the Franck-Condon transfer scheme requires the intermediate
state potential to be very deep and the initial and final state
potentials only moderately deep, therefore with a pure
spin-dependent optical dipole potential, there would be a high light
scattering rate since the dipole trap laser needs to be tuned
between D$1$ and D$2$ lines. As we shall see below, the transfer
time can be made of order of mili-seconds, thus if one uses
spin-dependent dipole potentials only for initial and final states
and keeps them on only during the transfer process, light scattering
will not be a serious problem. Furthermore, it is possible to create
blue-detuned spin-dependent optical potentials \cite{Das2006} which
could further reduce the light scattering in the initial and final
states.

\begin{figure}[h]
\includegraphics[width=0.95\linewidth]{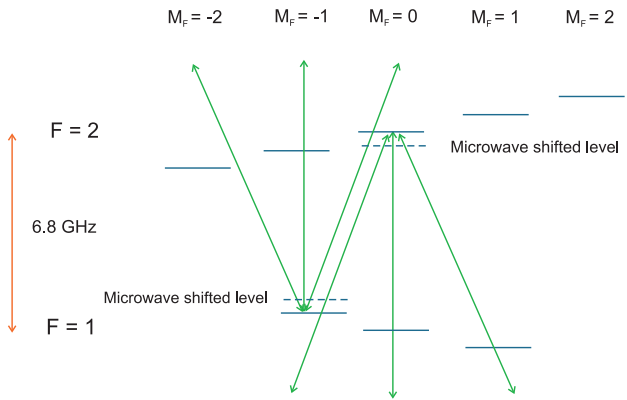}
\caption{\label{fig:microwave} Creating microwave potential for
$^{87}$Rb. For a near-field microwave positively detuned from both
$|1,-1\rangle\mapsto|2,0\rangle$ and $|1,1\rangle\mapsto|2,0\rangle$
transitions, one has a trapping potential for $|2,0\rangle$ and and
anti-trapping potentials for both $|1,-1\rangle$ and $|1,1\rangle$
(see Ref.~\cite{Treutlein2005}).}
\end{figure}

\begin{figure}[h]
\includegraphics[width=1.0\linewidth]{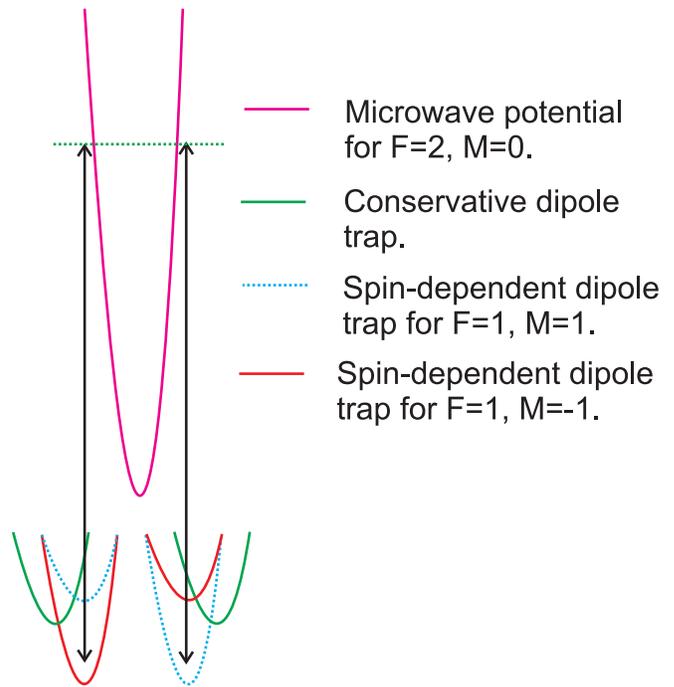}
\caption{\label{fig:microwave} One possible way of creating
spin-dependent potential profiles for three different states. The
net potential for atoms in initial and final states is the sum of
the one created by a conservative dipole trap and a spatially offset
spin-dependent dipole trap. The detuning and the polarization of the
spin-dependent dipole beams are chosen such that on the left hand
side, there is a good Franck-Condon overlap only for an atom in
$|F=1,m_{F}=-1\rangle$ state and on the right hand side, only for an
atom in $|F=1,m_{F}=1\rangle$ state. }
\end{figure}
%The transfer has to be driven by microwave transitions between
%different spin-states. It is interesting to note at this point that
%with microwaves, the Lamb-Dicke parameter $\eta$
%($=2\pi\times$oscillator length/wavelength) is essentially zero.
%Since the relevant matrix element for the transition between two
%vibrational states is $\langle
%n^{'}|e^{i\overrightarrow{k}.\overrightarrow{x}}|n \rangle
%$~\cite{Wineland1979}, if the potential profiles for two internal
%states are spatially the same (as in Raman sideband cooling
%experiments in Ion traps~\cite{Wineland1998}) then it is impossible
%to couple internal sates and vibrational states with $\eta = 0$.
%With spatially offset potential profiles as in our case, however,
%this is not a problem since the dipole approximation term is
%non-zero and with $\eta = 0$ the above matrix element is simply the
%Franck-Condon overlap. Physically, the sideways push on the atom
%comes from the different potential gradient experienced by the atom
%when transferred to a different internal state, so momentum is
%conserved. Such physical movements of a particle during internal
%state transition occurs also in the proposed scheme for microwave
%Raman sideband cooling of ions ~\cite{florian2001} where a
%differential magnetic field gradient is necessary for creating
%state-dependent potential profile.

To give an estimate of the parameter regime for which the spatial
transfer scheme works, we consider the situation of two harmonic
wells separated by a distance $1~\mu$m. The left well predominantly
traps atoms in $|F=1,m_{F}=-1\rangle$ state whereas the right well
traps predominantly $|F=1,m_{F}=1\rangle$ in the ground state of
$^{87}$Rb; states $|1\rangle$ and $|2\rangle$ respectively in the
earlier notation. This means that the left well is deep for atoms in
state $|1\rangle$ and shallow for atoms in $|2\rangle$, so that the
lowest vibrational wavefunction for atoms in $|1\rangle$ is much
more localized than that for atoms in $|2\rangle$, so that
Franck-Condon overlap for state $|2\rangle$ is negligible. Similarly
for right well, only overlaps for atoms in state $|2\rangle$ is
relevant. In between them is a tighter harmonic well with trap
frequency $\omega_{vib} = 2\pi\times50$ kHz which only traps a
$m_{F}$-state of the upper hyperfine level $F=2$, e.g.,
$|F=2,m_{F}=0\rangle$. The ground state of $^{87}$Rb has a hyperfine
splitting of $6.8$ GHz. The Zeeman energy is given by $g_Fm_F\mu_BB$
where $g_F = -1/4$ for F$=1$, so a bias magnetic field of a few
Gauss will create a situation similar to Fig.~\ref{fig:three}.
Spin-dependent potentials in the above-mentioned length scale and
trap frequencies are technically achievable using microwave
potentials as shown in~\cite{Treutlein2005}. For example, the
intermediate potential can be created by using near-field microwave
with a positive frequency detuning from both of $|F=1,m_{F}=1\rangle
\leftrightarrow |F=2,m_{F}=0\rangle$ and $|F=1,m_{F}=-1\rangle
\leftrightarrow |F=2,m_{F}=0\rangle$ transitions (Fig.3) so that one
has a trapping potential for an $F=2$ atom and anti-trapping
potential for an $F=1$ atom. Unlike in~\cite{Treutlein2005} where
the state $|F=2,m_{F}=1\rangle$ is used as a qubit state, in our
case the atom stays in the intermediate potential only for a very
short time (as we will see below) and therefore the choice of a
$F=2$ spin-state not crucial. The left and the right wells can be
created, for example, using a combination of conservative dipole
trap and a spin-dependent dipole trap with suitable choice of beam
polarization and detuning, so that the left trap provides good
Franck-Condon overlap only for atoms in $|F=1,m_{F}=-1\rangle$ state
and the right well, for $|F=1,m_{F}=1\rangle$. This can be further
ensured by adjusting the relative position and depth of the
conservative and spin-dependent dipole traps (Fig.4). The
spin-dependent part of the dipole trap can be switched off
adiabatically after the transfer to reduce spontaneous light
scattering. In the following, the overall trap frequencies of the
left and right wells for the appropriate spin-state is assumed to be
$2\pi\times10$ kHz. The spin-dependent dipole beam will also modify
the potential for the intermediate state equally on both sides and
this in fact enhances the overall Franck-Condon overlap. If we tune
the two microwaves connecting $|F=1,m_{F}=-1\rangle \rightarrow
|F=2,m_{F}=0\rangle$ and $|F=2,m_{F}=0\rangle \rightarrow
|F=1,m_{F}=1\rangle$ midway between the $13$th and $14$th
vibrational levels of the intermediate potential, the resultant of
the summation in Eq.($16$) is calculated to be $(0.053)/(2\pi\times
50)$kHz$^{-1}$. If the powers of the microwaves are adjusted so that
the Rabi frequencies $\widetilde{\Omega}_{3,1}$ and
$\widetilde{\Omega}_{3,2}$ in Eq.($16$) are $2\pi\times16$ kHz each
(the maximum of the Franck-Condon factors is $ \sim 0.2$, so the
individual two-level Rabi frequencies are $2\pi\times3.2$ kHz
$\ll\Delta$, since we assume $\Delta \sim 2\pi \times 25$ kHz,
condition ($10$) is satisfied. Impurity of transfer arising from
this approximation will be estimated below), the effective Rabi
frequency is about $2\pi\times 273$ Hz leading to a pi-pulse time
$\sim 1.8$ ms. The coherence time in microwave potentials is
estimated~\cite{Treutlein2005} to be of order of seconds, and up to
few tens of seconds in far-off resonant dipole traps, so the
transfer time is well-suited for coherent manipulations of cold
atoms for realizing phase-gates for quantum computation, inter alia.

It turns out that fairly large values of the summation is possible
even with small individual Franck-Condon factors. This arises from
the symmetry of the intermediate potential - the wavefunctions of
the alternate vibrational levels have opposite parities. Thus if we
tune the microwaves midway between the $n$ and $(n+1)$-th
vibrational levels, the matrix elements in the numerator of the term
inside the summation of Eq.($16$) have opposite signs for $n$ and
$(n+1)$, but since the detunings for $n$ and $(n+1)$ have opposite
signs too, the contributions from $n$ and $(n+1)$ add up. However,
for the same reason, the contribution from the pair $(n-1,n+2)$ will
tend to add destructively, but these levels are further detuned
(e.g. $\Delta_{n-1} = (1/3)\Delta_n$, for pure harmonic potential)
and overlap factors are also smaller. A great part of it is further
compensated by the contribution from the next pair $(n-2,n+3)$ Those
levels which do not have a counterpart have negligibly small
Franck-Condon factors and their contributions are not important.
This shows that it is helpful to choose the potential profiles such
that the Franck-Condon overlap falls off sharply above and below a
desired pair of vibrational levels. The parity of the wavefunctions
leading to alternate sign of the quantities in the summation also is
exactly the reason why the transfer is not possible in the optical
regime since when the vibrational levels cannot be resolved, the
contributions all cancel out (see Appendix).

It is also interesting note that in Eq.(17), the net detuning is the
Raman detuning as shown in Fig.2 minus the net AC Stark shift
generated by the Raman microwave beams summed over the vibrational
ladder of the intermediate potential. The AC Stark shift coming from
the individual beams
($(-1)^{p-1}\displaystyle\sum_{j}\Omega_{3j,p}/4\Delta_j$ with
$p=1,2$) is very small due to the alternating signs of $\Delta_j$
for levels $n$ and $n+1$ and similarly for all such pairs (as
discussed in the above paragraph) and given that the numerator was
always positive and in the numerical example given above, this is
less than $0.1$ percent of $\Omega_{eff}$. Of course, for symmetric
choice of Franck-Condon factors and Raman microwave beam powers
(i.e. $\Omega_{3j,1} = \Omega_{3j,2}$ for all $j$), the net AC Stark
shift (the term under summation in Eq.(17) is essentially zero.
Negligible AC Stark shifts for this Raman process is suitable for
the phase coherence of qubits required for quantum information
processing.

\begin{figure}[h]
\includegraphics[width=1.0\linewidth]{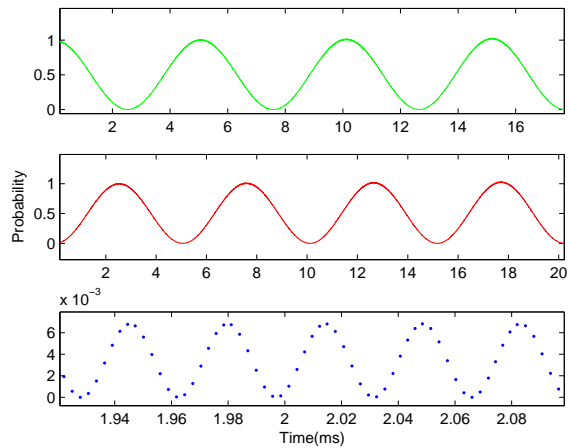}
\caption{\label{fig:adia}Population of the initial (top), final
(middle) and the intermediate (bottom) levels for the parameters
mentioned in the text.}
\end{figure}

To estimate the inaccuracy involved in the adiabatic elimination of
the intermediate levels, it is useful to consider the system as an
effective three-level system with an effective Franck-Condon factor
by re-writing Eq.($16$) in the following form:
\begin{eqnarray}
\Omega_{eff}&=&~\frac{\widetilde{\Omega}_{3,1}\widetilde{\Omega}_{3,2}}{2\Delta}{\bar{f}}_{eff},
\end{eqnarray}
Here $\Delta$ is half the mode-spacing of the levels $n$ and
$(n+1)$, midway between which the microwaves have been tuned and
$\bar{f}_{eff}$ is defined by comparing Eq.($19$) with Eq.(16). For
the simplified case where
$\widetilde{\Omega}_{3,1}\sqrt{{\bar{f}}_{eff}} =
\widetilde{\Omega}_{3,2}\sqrt{{\bar{f}}_{eff}} = \Omega$, the set of
equations ($7$), ($8$) and ($9$) become,
\begin{eqnarray}
 \frac{dc_1}{dt} &=& \frac{i\Omega}{2}e^{i\Delta t}c_3 \\
 \frac{dc_2}{dt} &=& \frac{i\Omega}{2}e^{i(\Delta - \delta) t}c_3\\
\frac{dc_3}{dt} &=& \frac{i\Omega}{2}e^{-i\Delta t}c_1 +
\frac{i\Omega}{2}e^{-i(\Delta - \delta) t}c_2
\end{eqnarray}
These coupled equations are numericalfly solved for the same
parameters as above and Fig.~\ref{fig:adia} shows the result. We can
see that the peak population of the intermediate level is below $1$
percent. This impurity decreases rapidly with increased pi-pulse
time (less power on the microwaves). It is interesting to note at
this point that with microwaves, the sideways push on the atom comes
from the different potential gradient experienced by the atom when
transferred to a different internal state, so momentum is conserved.
Such physical movements of a particle during internal state
transition occurs also in the proposed scheme for microwave Raman
sideband cooling of ions ~\cite{florian2001} where a differential
magnetic field gradient is necessary for creating state-dependent
potential profile.

\begin{figure}[h]
\includegraphics[width=0.75\linewidth]{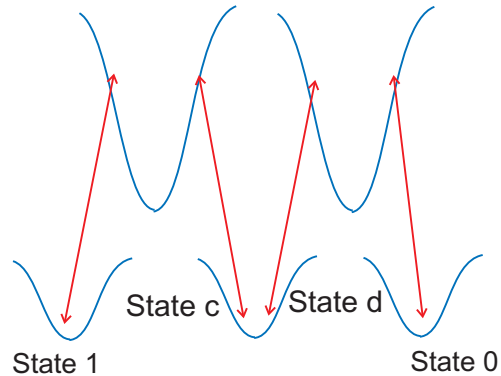}
\caption{\label{fig:phase_gate} The scheme for implementing a
collisional phase gate with Franck-Condon transfer. The states are
labelled $|1\rangle$ and $|0\rangle$ in the left and right wells
respectively according to the usual QIP conventions. Atoms starting
in state $|1\rangle$ are connected to the state $|c\rangle$ in the
central well, and similarly $|0\rangle \leftrightarrow |d\rangle$.}
\end{figure}

To illustrate how the spatial transfer scheme can be useful for
quantum information processing, let us consider the situation in
Fig.~\ref{fig:phase_gate}. Initially there is one atom in each of
the left and right wells. The intermediate potentials are designed
such that they provide suitable effective Franck-Condon factor only
for an atom in state $|1\rangle$ in the left well to transfer to
state $|c\rangle$ in the middle well and for an atom in state
$|0\rangle$ in the right well to transfer to state $|d\rangle$ in
the middle well. After a certain collisional time during which they
acquire a phase $\Phi$, they can be transferred back to their
initial wells and thus realizing a collisional phase gate
\cite{jaksch2005}
\begin{eqnarray*}
                          |1\rangle|1\rangle &=& |1\rangle|1\rangle \\
                         |1\rangle|0\rangle &=& |1\rangle|0\rangle e^{i\Phi} \\
                          |0\rangle|1\rangle &=& |0\rangle|1\rangle\\
                          |0\rangle|0\rangle &=& |0\rangle|0\rangle
                       \end{eqnarray*}
This will require two pairs of global pulses - one connecting state
$|1\rangle$ with state $|c\rangle$ and the other connecting
$|0\rangle$ with $|d\rangle$. In the context of $^{87}$Rb, states
$|c\rangle$ and $|d\rangle$ can both be $|F=1, m_F = 0\rangle$ and
$|0\rangle$ and $|1\rangle$ can be $|F=1, m_F = -1\rangle$ and
$|F=1, m_F = 1\rangle$ respectively. In the case where both atoms in
left well and right well go to the central well, separating them
afterwards at the end of the phase accumulation can be a problem,
which can, at least in principle, be surmounted by using the
adiabatic transfer technique described in the next paragraph. Using
other atomic species with a wider choice of spin-states, one can
have states $|c\rangle$ and $|d\rangle$ different to each other (and
to $|0\rangle$ and $|1\rangle$), e.g. for $^{133}$Cs the lowest
ground hyperfine state ($F=3$) has seven different spin-states.

One of the preliminary requirements for using neutral atoms in
lattice structures as scalable quantum computers~\cite{jaksch1999}
is to deterministically prepare one atom per lattice site in the
motional ground state. Several proposals to this end have been made
~\cite{rabl2003},~\cite{diener2002},~\cite{mohring2005}. In the
following we show how the spatial separation scheme can be useful
for realizing the purification scheme suggested in~\cite{rabl2003}
which consists of an adiabatic transfer of an atom from one internal
state with low inter-atom interaction (Hubbard-U) to another with a
high interaction energy (e.g., by choosing the second internal state
to be one with a Feshbach resonance and applying a magnetic field
close to the resonance to get the required high U). Initially, a
small number of atoms ($N$ $=~3,4$, on an average) is loaded into
the first internal state from a BEC with a Poissonian number
distribution, so the number uncertainty is $\sim\sqrt{N}$ leading to
an uncertainty in the interaction energy of the atoms in the first
internal state $\sim \sqrt{N}U$, and a very low probability of there
being zero atoms. Adiabatically scanning the Raman detuning over a
suitable range can ensure that only one atom is transferred to the
second internal state in the same well. State-selectively removing
the remaining atoms in the first internal state leaves one with a
pure atomic crystal with one atom in the motional ground state of
each site. A scheme in which atoms are transferred between internal
states \emph{within} the same well means that there must be a
separation of these states at the end of the process. Removing the
atoms state-dependently while maintaining high purity of the desired
single-atom state is not straightforward: an obvious option is to
use a pulse of resonant light to kick the unwanted atoms out of the
trap. However if these atoms are in the same potential well,
inelastic light scattering can leave some of the atoms in the second
state which limits the purity of the final state. Using a
spin-dependent lattice to separate the two spin-species spatially is
another option but for states which can be connected by magnetic
dipole transitions, it is not possible to make the force on one
spin-species more than 3 times bigger than that on the other at low
magnetic fields, so one needs to effect the separation process very
slowly. Another possibility is to transfer the atoms into the first
excited Bloch band of the lattice and then lower the depth of the
lattice such that the atoms in the first internal state are released
while the only atom in the second internal state is still
trapped~\cite{rabl_pri}, but for a number of closely spaced wells
this may lead to tunneling, and thus affect the purity. Also in all
these schemes, atoms in their initial and final internal states are
in the same region of space and thus collisional loss such as
three-body recombination will limit the purity. With the
Franck-Condon spatial transfer scheme, since the atom transferred to
the second state is in a different well, removing the atoms in the
first internal state is straightforward and there is no collisional
loss. Suppose we want to transfer only one atom to the right well in
Fig.~\ref{fig:three}, starting with a few atoms (N $\sim4$) in the
left well with a number uncertainty of $\sim \sqrt{N}$. For this
scheme, it is advantageous to have less tight confinement in the
starting well and strong confinement in the final well to make the
ratio of on-site interaction energy in the final and initial wells
as large as possible. Such adjustments are possible whilst keeping
good overlap of wavefunctions. This means the region of desirable
avoided crossings between number states (i.e. ones between
$|N,0\rangle \rightarrow |N-1,1\rangle$ with a $\sqrt{N}$
uncertainty in N) is small enough to be scanned adiabatically, and
this region is small enough so that condition ($14$) is fulfilled
and the region of unwanted avoided crossings ($|N-1,1\rangle
\rightarrow |N-2,2\rangle$) are driven far from the former region.
For $N_{ave} = 4$, trap frequencies for left and right well
$=2\pi\times 3$kHz and $=2\pi\times 15$kHz, same bare Rabi
frequencies and $\Delta$ as above and Raman detuning and Rabi
frequency ramps chosen as suggested in~\cite{rabl_thesis}, a
transfer time of $35$ ms gives $99.5\%$ purity of transfer. As
suggested in ~\cite{rabl2003}, the transfer time can be minimized by
optimizing pulse shapes for Raman detuning and effective Rabi
frequency. Switching off the traps holding the atoms in the initial
state will leave an array of wells, each of which contains only a
single atom.

%One can stretch the initial wavepacket at the cost of squeezing the
%final wavepacket and keep the effective Franck-Condon factor and
%effective Rabi frequency the same and at the same time making the
%ratio of on-site interaction energy of the atom in the final and
%initial states big, so that transition of more than one atom is
%suppressed.

In summary, we have discussed a scheme of spatially transferring
atomic wavepackets between different micro-wells and derived the
conditions for such a transfer scheme to work effectively. We have
shown that such a transfer can work in an experimentally achievable
parameter regime and can be useful for realizing two-qubit phase
gates and realizing a deterministic single atom occupation scheme on
a realistic time scale. Such a scheme can also be useful for atomic
interferometry with localized atomic wavepackets.

\begin{section}{Appendix}
 In this appendix we show why the Franck-Condon transfer cannot easily be implemented using transitions between
 electronic states (which would be the most direct analogy with molecular transitions such as that shown in Fig.1b).
  Using $^{87}$Rb, for example, a state-dependent
potential profile can be created by using two focussed Gaussian
beams red-detuned from the $5^{2}S_{1/2}$ - $5^{2}P_{3/2}$
transition to form the initial and the final wells and inserting in
between the beams a central well formed by another focussed Gaussian
beam red-detuned from, say, $5^{2}P_{3/2}$ - $6^{2}P_{3/2}$. The
problem with this kind of potential for the current scheme is that
the natural linewidths of such transitions are several MHz or more.
To make the spacing of the vibrational states of the intermediate
states many MHz so that they could be resolved, one needs the
intermediate potential very deep which is practically impossible for
any available laser power and without enormous loss due to
spontaneous emission. When the vibrational states cannot be
resolved, $\Delta_{n}$ in the denominator of Eq.($16$) is same for
all $n$:
\begin{eqnarray*}
    \Omega_{eff} &=& \frac{\widetilde{\Omega}_{3,1}\widetilde{\Omega}_{3,2}}{2\Delta}\displaystyle\sum_{n}\langle n_1=1|n\rangle\langle n|n_2=1\rangle \\
    &=&\frac{\widetilde{\Omega}_{3,1}\widetilde{\Omega}_{3,2}}{2\Delta}\langle n_1=1|n_2=1\rangle,
    \end{eqnarray*}
since by completeness,
\begin{equation*}
    \displaystyle\sum_{n}|n\rangle\langle
    n| = 1.
\end{equation*}
The direct overlap of the two lowest vibrational levels of the two
spatially separated traps is essentially very small. Thus the
Franck-Condon (FC) transfer scheme cannot be carried out using
optical transitions.

\end{section}

\vspace{0.65in}

\begin{acknowledgments}
We acknowledge support from EPSRC, EC (Marie-Curie fellowship, Cold
Quantum Gases network), St. John's College, the Royal Society, and
DARPA.
\end{acknowledgments}

\end{document}